\documentclass[grl]{agutex2arxiv}

\usepackage{siunitx}
\usepackage{graphicx,xcolor,amsmath,amssymb}
\usepackage{lineno}

\pdfoutput=1
\setkeys{Gin}{draft=false}

\authorrunninghead{DAVID ET AL.}
\titlerunninghead{ATTENUATION IN ANTIGORITE}

\authoraddr{E.C. David,
  Department of Earth Sciences,
  University College London, Gower Street, London WC1E 6BT, UK.
  (e.david@ucl.ac.uk)}

\begin{document}

\title{Low-Frequency Measurements of Seismic Velocity and Attenuation in Antigorite Serpentinite}

\authors{Emmanuel C. David\altaffilmark{1},
 Nicolas Brantut\altaffilmark{1}, Lars N. Hansen\altaffilmark{2}, and Ian Jackson\altaffilmark{3}}

\altaffiltext{1}{Department of Earth Sciences, University College London, London, United Kingdom.}
\altaffiltext{2}{Department of Earth Sciences, University of Oxford, Oxford, United Kingdom.}
\altaffiltext{3}{Research School of Earth Sciences, Australian National University, Canberra, Australia.}

%\begin{keypoints}
%\item	Attenuation in antigorite increases noticeably with increasing oscillation period and temperature above \SI{500}{\celsius}.
%\item	Viscoelastic relaxation in antigorite is described by a Burgers model and possibly results from intergranular mechanisms.
%\item Attenuation is higher in antigorite than in olivine, but such contrast is much less than the contrast in shear velocity.
%\end{keypoints}

\begin{abstract}
Laboratory measurements of seismic velocity and attenuation in antigorite serpentinite at a confining pressure of \SI{2}{\kilo\bar} and temperatures up to \SI{550}{\celsius} (\textit{i.e.}, in the antigorite stability field) provide new results relevant to the interpretation of geophysical data in subduction zones. A polycrystalline antigorite specimen was tested via forced-oscillations at small strain amplitudes and seismic frequencies (\si{\milli\hertz}--\si{\hertz}). The shear modulus has a temperature sensitivity, $\partial G/ \partial T$, averaging \SI{-0.017}{\giga\pascal\per\kelvin}. Increasing temperature above \SI{500}{\celsius} results in more intensive shear attenuation ($Q_G^{-1}$) and associated modulus dispersion, with $Q_G^{-1}$ increasing monotonically with increasing oscillation period and temperature. This ``background'' relaxation is adequately captured by a Burgers model for viscoelasticity and possibly results from intergranular mechanisms. Attenuation is higher in antigorite ($\log_{10} Q_G^{-1} \approx -1.5$ at \SI{550}{\celsius} and \SI{0.01}{\hertz}) than in olivine ($\log_{10} Q_G^{-1} \ll -2.0$ below \SI{800}{\celsius}), but such contrast does not appear to be strong enough to allow robust identification of antigorite from seismic models of attenuation only. 
\end{abstract}

\begin{article}
  
\section{Introduction}
\label{sec:intro}

The interpretation of geophysical data and the associated need to provide constraints on the presence of serpentinites, notably in subduction zones, has motivated a number of laboratory studies reporting measurements of P- and S-wave velocities under controlled conditions of pressure and temperature, on both single-crystal \citep{Bezacieretal2010,Bezacieretal2013} and polycrystalline serpentinites \citep[see \textit{e.g.}][]{Birch1960,Christensen1978,Kernetal1997,Watanabeetal2007,Jietal2013,Davidetal2018}. However, geophysical observations of \emph{seismic attenuation} (or its inverse, the quality factor $Q$) in subduction zones \citep[see \textit{e.g.}][]{Eberhardetal2008,P09,Wangetal2017} have generally defied conclusive interpretation because of the crucial lack of directly relevant laboratory data for serpentinites. Correlations between low-$Q$ anomalies and serpentinization are usually based on joint information from velocity tomography and the commonly held view that the presence of serpentinites is associated with low velocities and high Poisson's ratio \citep{HyndmanPeacock2003}. Ultrasonic velocity measurements demonstrate that while this correlation might be appropriate for lizardite or chrysotile serpentinites, it is not valid for antigorite serpentinites \citep{Christensen2004,Reynard2013,Davidetal2018}. In addition, seismic attenuation (particularly shear attenuation) is very sensitive to many factors, including temperature, grain size, dislocation density, melt fraction and redox conditions \citep[see \textit{e.g.}][]{Jackson2015,Clineetal2018}.

Among the serpentine group, antigorite is the mineral stable over the widest depth range \citep{UlmerTrommsdorff1995}. The only existing, published attenuation data in serpentinites were reported on an anisotropic antigorite serpentinite by \citet{Kernetal1997}, who measured the directional dependence of ultrasonic P- and S-wave velocities and their associated attenuations up to \SI{6}{\kilo\bar} confining pressure and \SI{700}{\celsius}. More recently, \citet{Sviteketal2017} performed similar measurements for P-waves up to \SI{4}{\kilo\bar} confining pressure. However, such attenuation (and velocity) data were only obtained at the single MHz frequency of the ultrasonic pulse transmission method, which limits seismological application of the results as most rocks exhibit solid-state viscoelastic relaxation at lower frequencies, notably when temperature is increased \citep{Jackson2015}.

This lack of data motivated the present laboratory study for measuring shear modulus and attenuation on an isotropic antigorite specimen. The use of forced-oscillation techniques allows for mechanical spectroscopy in the \si{\milli\hertz}--\si{\hertz} frequency range and under low strain amplitudes ($<10^{-5}$) analogous to those of seismic wave propagation. Measurements were carried out in the ``Jackson-Paterson'' gas medium apparatus \citep{JacksonPaterson1993}, from room temperature to \SI{550}{\celsius}, in the antigorite stability field. The applied confining pressure of \SI{2}{\kilo\bar} is expected to be sufficient to suppress most of the contribution from open microcracks (as demonstrated by a recent study on the same material by \citet{Davidetal2018}) or from other effects arising from frictional sliding on crack surfaces and grain boundaries \citep{Mavko1979,Jacksonetal1992}. Additional measurements under flexural forced oscillation, providing access to Young's modulus and associated attenuation, are also reported at selected temperatures. Discussion of the results focuses on the temperature-dependence of shear modulus, on the possible mechanisms responsible for the observed viscoelastic relaxation, and on a comparison with existing data on olivine followed by concluding remarks on the geophysical significance of the new laboratory data.

\section{Experimental Materials and Methods}
\label{sec:exp}

\begin{figure*}
  \centering
\includegraphics[width=\textwidth]{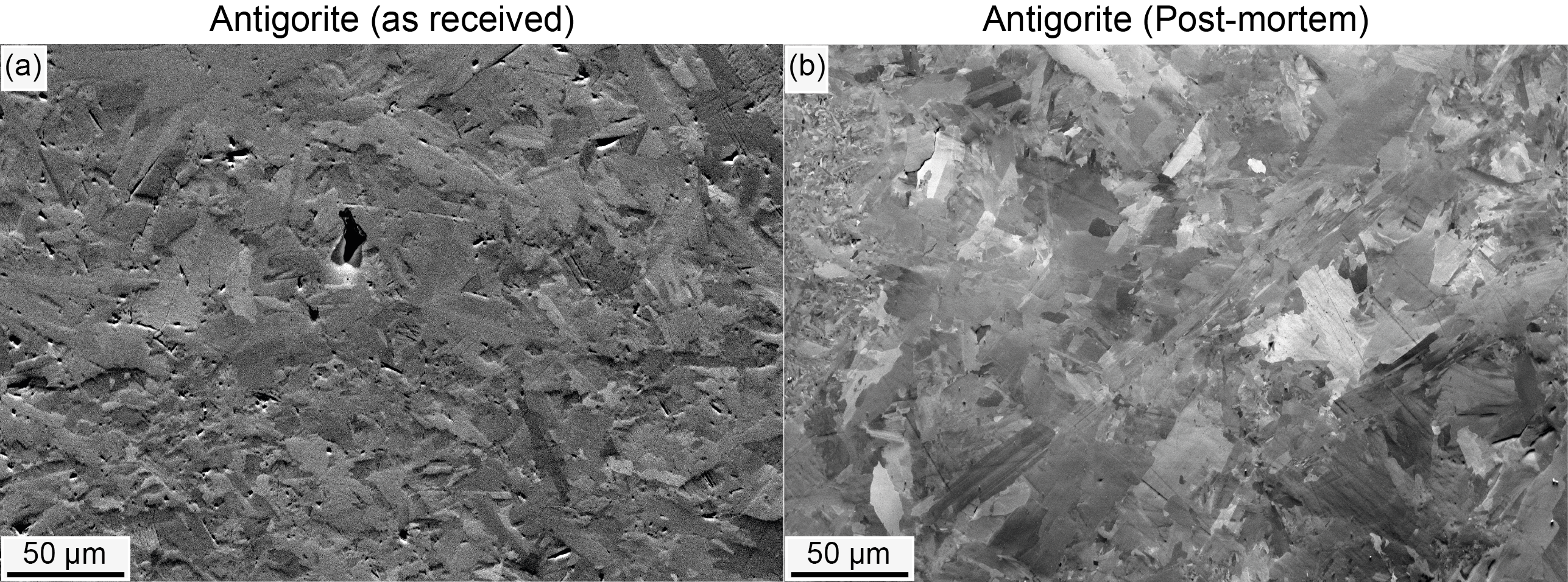}
\caption{Forescatter electron images of antigorite. a) Specimen as received (after \citet{Davidetal2018}); b) Transverse section of specimen recovered post-mortem after exposure to a maximum temperature of \SI{550}{\celsius}. The forescatter electron images were acquired using the Oxford Instruments AZtec software on an FEI Quanta 650 field-emission gun scanning-electron microscope in the Department of Earth Sciences at the University of Oxford.}
  \label{fig:micro}
\end{figure*}

A specimen was cored from a block of ``Vermont antigorite serpentinite'' acquired from Vermont Verde Antique's Rochester quarry (Vermont, USA). This block is the same material as the ``isotropic'' block described in detail in \citet{Davidetal2018}. The rock is nearly pure antigorite ($>$~95$\%$), with minor amounts of magnetite and magnesite (both at about 2$\%$ level). The antigorite grains are extremely heterogeneous in both size and shape (Figure~\ref{fig:micro}a). The material is mainly fine-grained, but grain size ranges from sub-micron to a few hundreds of microns. The antigorite grains are generally elongated with random orientations of their long axes. The rock is elastically isotropic, with P- and S-wave velocities averaging \num{6.5} and \SI{3.7}{\kilo\metre\per\second}, respectively, at room pressure and temperature. Rock density is \SI{2.65e3}{\kilogram\per\cubic\metre}.

The cylindrical core specimen was then precision ground to \SI{11.50}{\milli\metre} in diameter and \SI{31.14}{\milli\metre} in length and clamped between a series of optically flat Lucalox$^\textrm{TM}$ alumina pistons (see Figure 1 of \citet{Jacksonetal2009}). This assembly was enclosed within a thin-walled (\SI{0.35}{\milli\metre}) copper jacket and loaded into the ``Jackson-Paterson'' gas apparatus at the Australian National University (Canberra, Australia) for high-temperature mechanical testing using low-frequency forced oscillations \citep{JacksonPaterson1993}. The specimen was pressurised to a confining pressure of \SI{2}{\kilo\bar} for the entire experiment. Temperature was then raised in \SI{50}{\celsius} intervals to the maximum targeted temperature of \SI{550}{\celsius}. This maximum temperature was chosen to conservatively cover most of the antigorite stability field while avoiding dehydration \citep{UlmerTrommsdorff1995}. The specimen was annealed for \SI{2}{\hour} at \SI{550}{\celsius} and then staged-cooled in \SI{50}{\celsius} intervals down to room temperature using the same ramp as for staged-heating (\SI{450}{\celsius\per\hour}). After completion of the experiment, examination of microstructures in the sectioned antigorite specimen (Figure~\ref{fig:micro}b) indicates that the staged heating, exposure to annealing at \SI{550}{\celsius} and subsequent staged cooling have resulted in no measurable microstructural evolution relative to the starting material (Figure~\ref{fig:micro}a), with no evidence of dehydration.

\emph{Torsional} forced-oscillation tests were conducted at a series of ten logarithmically equispaced periods between 1 and 1000~\si{\second}, at selected temperatures, during \emph{both} staged heating and staged cooling. It was observed that the mechanical behaviour during staged heating and cooling was reproducible, which is consistent with the absence of microstructural evolution during the entire experiment. Here we only report data obtained during \emph{staged cooling}, as we consider those to be the most representative of the stable microstructure. Several forced-oscillation tests were performed at \SI{500}{\celsius} to check for both temporal evolution and amplitude dependence of the mechanical behaviour (Figures~\ref{fig:1672GQlinsens}a and \ref{fig:1672GQlinsens}b). The reason for conducting such tests at \SI{500}{\celsius}, rather than at the maximum temperature of \SI{550}{\celsius}, was again to avoid any dehydration in the antigorite specimen. The mechanical behaviour was unaffected by increasing exposure time (Figures~\ref{fig:1672GQlinsens}a and \ref{fig:1672GQlinsens}b), a result that is again consistent with the absence of microstructural evolution in the antigorite specimen. However, the mechanical response (particularly the shear modulus) was amplitude-dependent for applied shear stresses above \SI{1200}{\kilo\pascal} (Figures~\ref{fig:1672GQlinsens}a and \ref{fig:1672GQlinsens}b), corresponding to maximum shear strains $\gtrsim 1.5 \times 10^{-5}$  in the antigorite specimen. Accordingly, in order to maximise the signal to noise ratio whilst satisfying the condition of linear viscoelastic behaviour, a shear stress amplitude of \SI{1200}{\kilo\pascal} was used during staged cooling to room temperature. 

At a given temperature and oscillation period, a sinusoidally time-varying torque was applied by a pair of electromagnetic drivers located at the lower end of the assembly (see \textit{e.g.} Figure 1a of \citet{ClineJackson2016}). Time-varying displacements were measured by pairs of parallel-plate capacitance transducers mounted off-axis for mechanical advantage. Examples of raw data are given in \citet{JacksonPaterson1993}. The procedure for estimation of shear modulus ($G$) and attenuation ($Q_G^{-1}$) from torsional forced oscillation data required a parallel series of experiments to be conducted on a ``reference assembly'' in which the antigorite specimen is replaced by a single-crystal sapphire specimen, as described in \citet{JacksonPaterson1993}. Importantly, such procedure also involved a correction for the viscoelasticity of the copper jacket. Extensive data on small-strain viscoelasticity of copper were recently reported by \citet{DavidJackson2018} for copper annealed at \SI{1050}{\celsius} and \SI{900}{\celsius}. It was observed that although the different annealing conditions result in substantially different grain growth in copper, there is only a modest difference in the viscoelastic behaviour. Unfortunately, no equivalent data are available for copper annealed at \SI{550}{\celsius}. Hence, the sensitivity of the mechanical behaviour to the jacket correction was tested by comparing results obtained using the viscoelastic properties of copper annealed at 1050 and 900~\si{\celsius}, at selected temperatures (Figures~\ref{fig:1672GQlinsens}c and \ref{fig:1672GQlinsens}d). The two distinct jacket corrections did not noticeably affect the period-dependence of shear modulus and attenuation (for $\log_{\mathrm{10}} (Q_G^{-1}) \gtrsim 2.2$), and only resulted in a small offset of about 1--2~\si{\giga\pascal} for the shear modulus, which is a direct estimate of the accuracy of measurements. Accordingly, because of a denser sampling in the temperature interval $\leq$\SI{550}{\celsius}, the data of \citet{DavidJackson2018} on the viscoelasticity of copper annealed to \SI{1050}{\celsius} were used for the jacket correction in this study.

In addition to torsional oscillations, \emph{flexural} forced-oscillation tests were conducted in the same frequency range as for torsional oscillations at selected temperatures during staged heating (\num{20}, \num{300} and \SI{500}{\celsius}) by using a sinusoidally time-varying bending force and an appropriate arrangement of the capacitance transducers for measuring bending displacements (see, \textit{e.g.}, Figure 1c of \citet{ClineJackson2016}). The procedure for extraction of Young's modulus ($E$) and its associated attenuation ($Q_E^{-1}$) from flexural forced-oscillation data involved forward modelling with a finite-difference filament elongation model for flexure of the long thin experimental assembly. This procedure is described in \citet{Jacksonetal2011} and \citet{ClineJackson2016}.

The variation of both shear modulus and attenuation as functions of oscillation period and temperature can be fitted by a model for linear viscoelasticity based on a Burgers-type creep function \citep{JacksonFaul2010}. The specific features of the Burgers model used in this study are recalled here. The model incorporates a broad distribution $D_{\scriptscriptstyle \mathrm{B}}(\tau)$ of anelastic relaxation times, $\tau$, to account for the monotonic ``background'' dissipation and associated modulus dispersion as:
\begin{linenomath*}
\begin{equation}
  \label{eq:anelabroad}
D_{\scriptscriptstyle \mathrm{B}}(\tau)=\frac{\alpha \tau^{\alpha-1}}{\tau_{\scriptscriptstyle \mathrm{M}}^{\alpha}-\tau_{\scriptscriptstyle \mathrm{L}}^{\alpha}}
\end{equation}
\end{linenomath*} 
with $0<\alpha<1$ for $\tau_{\scriptscriptstyle \scriptscriptstyle \mathrm{L}}<\alpha<\tau_{\scriptscriptstyle \scriptscriptstyle \mathrm{M}}$ and zero elsewhere \citep{MinsterAnderson1981}. In equation \eqref{eq:anelabroad}, the upper limit $\tau_{\scriptscriptstyle \mathrm{M}}$ is identified with the characteristic time  for Maxwell relaxation, giving way to \emph{viscous relaxation} at sufficiently long timescales ($\gg \tau_{\scriptscriptstyle \mathrm{M}}$). The variation of $G$ and $Q_G^{-1}$ with angular frequency $\omega=2\pi/T_0$ is prescribed by
\begin{linenomath*}
\begin{equation}
    \label{eq:GQBurg}
    G(\omega)=[J_1^2(\omega)+J_2^2(\omega)]^{-1/2} \qquad \textrm{and} \qquad Q_G^{-1}=J_2(\omega) / J_1(\omega),
\end{equation}
\end{linenomath*}
where ($J_1(\omega)$,$J_2(\omega)$) are, respectively, the real and negative imaginary parts of the complex compliance $J^{\star}(\omega)$, which are explicitly related to the distribution of relaxation times as follows \citep{JacksonFaul2010}
\begin{linenomath*}
\begin{eqnarray}
  J_1(\omega) & = & \frac{1}{G_{\scriptscriptstyle \mathrm{U}}} \Big[1+\Delta_{\scriptscriptstyle \mathrm{B}} \int_{\tau_{\scriptscriptstyle \mathrm{L}}}^{\tau_{\scriptscriptstyle \mathrm{M}}} \frac{D_{\scriptscriptstyle \mathrm{B}}(\tau)}{1+\omega^2 \tau^2} d\tau \Big] \label{eq:J1},\\
  J_2(\omega) & = & \frac{1}{G_{\scriptscriptstyle \mathrm{U}}}\Big[\omega \Delta_{\scriptscriptstyle \mathrm{B}} \int_{\tau_{\scriptscriptstyle \mathrm{L}}}^{\tau_{\scriptscriptstyle \mathrm{M}}} \frac{\tau D_{\scriptscriptstyle \mathrm{B}}(\tau)}{1+\omega^2 \tau^2} d\tau + \frac{1}{\omega\tau_{\scriptscriptstyle \mathrm{M}}} \Big] \label{eq:J2},
\end{eqnarray}
\end{linenomath*}
where $G_{\scriptscriptstyle \mathrm{U}}$ is the unrelaxed shear modulus and $\Delta_{\scriptscriptstyle \mathrm{B}}$ is the relaxation strength.

The \emph{temperature dependence} of each of the two characteristic relaxation times ($\tau_{\scriptscriptstyle \mathrm{L}}$,$\tau_{\scriptscriptstyle \mathrm{M}}$) relative to their respective values ($\tau_{\scriptscriptstyle \mathrm{LR}}$,$\tau_{\scriptscriptstyle \mathrm{MR}}$) at reference temperature $T_{\scriptscriptstyle \mathrm{R}}$ follows the Arrhenian expression
\begin{linenomath*}
\begin{equation}
  \label{eq:Arrhenius}
  \tau_i(T)=\tau_{iR} \Big[e^{\displaystyle (E_{\scriptscriptstyle \mathrm{B}}/R)(1/T-1/T_{\scriptscriptstyle \mathrm{R}})}\Big],
\end{equation}
\end{linenomath*}
where $i=\mathrm{L}$ or $i=\mathrm{M}$ for the broad anelastic relaxation background, associated with an activation energy $E_{\scriptscriptstyle \mathrm{B}}$; $R$ is the gas constant. Finally, the temperature dependence of the unrelaxed shear modulus, $G_{\scriptscriptstyle \mathrm{U}}$, relative to its value at the reference temperature, $G_{\scriptscriptstyle \mathrm{U}}(T_{\scriptscriptstyle \mathrm{R}})$, is simply specified as
\begin{linenomath*}
\begin{equation}
  \label{eq:Gu}
  G_{\scriptscriptstyle \mathrm{U}}(T)=G_{\scriptscriptstyle \mathrm{U}}(T_{\scriptscriptstyle \mathrm{R}})+(T-T_{\scriptscriptstyle \mathrm{R}})(dG_{\scriptscriptstyle \mathrm{U}}/dT_{\scriptscriptstyle \mathrm{R}}).
\end{equation}
\end{linenomath*}
%The list of fitting parameters of the Burgers model is given in Table~\ref{tab:burg}.

\section{Results}
\label{sec:res}

\begin{figure*}
  \centering
\includegraphics[width=\textwidth]{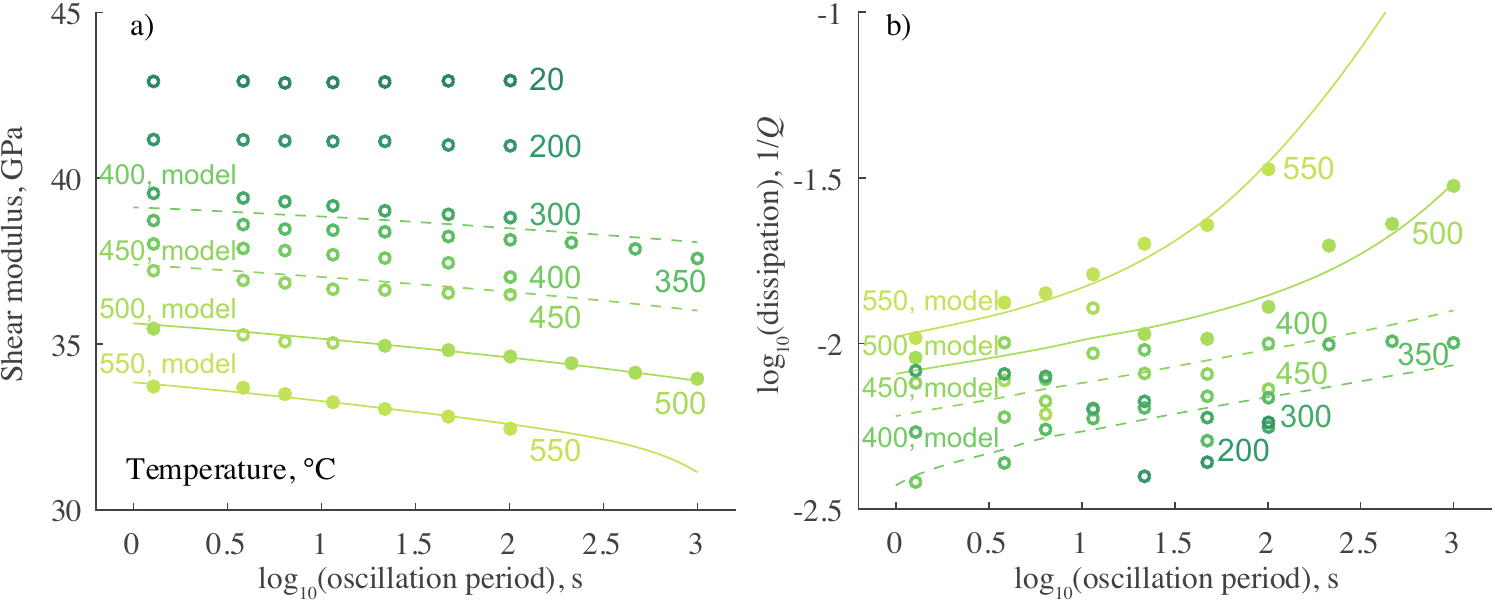}
\caption{Shear modulus and dissipation data (symbols) and optimal Burgers model (curves) for antigorite. The scatter in the dissipation data provides an indication of experimental uncertainties (\textit{n.b.} at room temperature, $\log_{10} Q_G^{-1}$~$<$~-2.5 at all oscillation periods). The Burgers model (see Table~\ref{tab:burg}) was fitted to $N$~=~14 $(G,Q_G^{-1})$ data pairs (full symbols) in the \num{500}--\SI{550}{\celsius} temperature range and extrapolated to \num{450} and \SI{400}{\celsius} (dashed curves).}
  \label{fig:1672GQ}
\end{figure*}

Experimental results for the variation of shear modulus and attenuation with oscillation period and temperature in the antigorite specimen are presented in Figure~\ref{fig:1672GQ}. Note that the reason for not conducting forced-oscillation tests at longer oscillation periods ($>$ \SI{101}{\second}) at all temperatures was that shear modulus and dissipation were not found to be strongly frequency-dependent below \SI{450}{\celsius}.

The shear modulus decreases systematically with increasing temperature and, at all temperatures above \SI{300}{\celsius}, with increasing oscillation period (Figure~\ref{fig:1672GQ}a). The strength in dispersion of the shear modulus increases slightly with temperature to reach about 4$\%$ at \SI{550}{\celsius} across the 1--100~\si{\second} oscillation period interval. Below \SI{450}{\celsius}, $\log_{10} Q_G^{-1} < -2$ and is essentially period-independent, with a tendency to increase slightly with temperature (Figure~\ref{fig:1672GQ}b); the scatter in the data provides an indication of experimental uncertainties. Increasing temperature above \SI{500}{\celsius} results in more intensive shear attenuation, with $Q_G^{-1}$ increasing noticeably with temperature and oscillation period. Over the observational oscillation period window, the monotonic increase of $Q_G^{-1}$ with oscillation period $T_0$ above \SI{500}{\celsius} is typical of the ``absorption band'' commonly observed in many rocks at elevated temperatures \citep{Cooper2002,MinsterAnderson1981,Jacksonetal1992} and adequately approximated by a power law of the form $Q_G^{-1} \sim T_0^{\alpha}$ with $\alpha$~=~0.20 and $\alpha$~=~0.26 at \SI{500}{\celsius} and \SI{550}{\celsius}, respectively.

The variation of both shear modulus and attenuation with oscillation period and temperature in the \num{400}--\SI{550}{\celsius} range is reasonably well described by a ``background relaxation only'' Burgers-type model for linear viscoelasticity (Figure~\ref{fig:1672GQ}; fitting parameters in Table~\ref{tab:burg}). The model was fitted to the $(G,Q_G^{-1})$ data pairs for which $Q_G^{-1}$ is systematically period-dependent (\textit{i.e.}, at 500--550\si{\celsius}) and extrapolated to lower temperatures. That shear modulus and dissipation data are jointly fitted by the Burgers model, based on a creep function and ensuring compliance with the linear Kramers-Kronig relations, is confirmation that the measurements were taken in the regime of linear viscoelasticity. The trends in variation of both shear modulus and dissipation with oscillation period are well captured by the model in the \num{400}--\SI{550}{\celsius} range, but the temperature-dependence of the shear modulus becomes increasingly offset at \SI{400}{\celsius}. Part of the misfit is attributed to the relatively sharp increase in the dissipation and associated modulus dispersion data which occurs between \num{450} and \SI{500}{\celsius}. The characteristic time for Maxwell relaxation is $\log_{10}(\tau_{\mathrm{M}})$~=~-1.7 at \SI{500}{\celsius} (Table~\ref{tab:burg}) and $\log_{10}(\tau_{\mathrm{M}})$~=~-0.6 at \SI{550}{\celsius} (Table~\ref{tab:burg} and equation~\eqref{eq:Arrhenius}), which possibly indicates a noticeable contribution of viscous relaxation to the dissipation observed at these temperatures.

\begin{table}[t]
  \caption{Values for the 7 parameters of the optimal Burgers model for antigorite (see Figure~\ref{fig:1672GQ}), fitted to $N$~=~14 ($G$,$Q_G^{-1}$) data pairs using a reference temperature $T_{\scriptscriptstyle \mathrm{R}}$~=~\SI{500}{\celsius}. Model misfit is $\sqrt{(\chi_G^2+\chi_{Q^{-1}_G}^2)/(2N)}$~=~0.55 with \textit{a priori} errors of $\sigma(G)/G$~=~0.03 and $\sigma[\log_{10}(Q_G^{-1})]$~=~0.05. Parameter uncertainties are parenthesised after parameter value, and parameters that are bracketed indicate that value was kept fixed for purposes of convergence in the fitting strategy.}
  \label{tab:burg}
  \centering
  \begin{tabular}{|c|c|}
    \hline
    $G_{\scriptscriptstyle \mathrm{U}}$, \si{\giga\pascal} & 36.6(0.5) \\
    $dG_{\scriptscriptstyle \mathrm{U}}/dT$, \si{\giga\pascal\per\kelvin} & -0.026(0.012) \\
    $\Delta_{\scriptscriptstyle \mathrm{B}}$ & 0.13(0.01) \\
    $\alpha$ & 0.10(0.03) \\
    $\log_{10}(\tau_{\scriptscriptstyle \mathrm{LR}})$, \si{\second} & [-1.7] \\
    $\log_{10}(\tau_{\scriptscriptstyle \mathrm{MR}})$, \si{\second} & 1.7(0.05) \\
    $E_{\scriptscriptstyle \mathrm{B}}$, \si{\kilo\joule\per\mole} & 268(21) \\
    \hline
  \end{tabular}
  %}
\end{table}

Flexural forced-oscillation data, in the form of ``normalised flexural compliance'' and ``specimen assembly phase lag'' obtained at representative temperatures, are displayed in Figure~\ref{fig:1672flex}. Although the flexural forced-oscillation data (particularly the specimen assembly phase lag data) are significantly more scattered than their torsional counterparts, it is clear that the normalised flexural compliance increases with increasing temperature between room temperature and \SI{500}{\celsius} (Figure~\ref{fig:1672flex}a). The data are too scattered to provide any resolvable period-dependence for either the normalised flexural compliance and specimen assembly phase lag data. The flexural oscillation data are compared with the finite-difference approximation to the filament elongation model in Figure~\ref{fig:1672flex}. Importantly, the model relies on the assumption that the Young's modulus, at each location along the temperature gradient, is prescribed by combining the complex shear modulus for specimen or alumina connecting rods and the copper jacket, previously determined in torsional forced-oscillation tests, with the corresponding \emph{anharmonic temperature-dependent bulk modulus}. That the model predictions are in semi-quantitative accord with the observations indicates that such assumption is reasonable for the antigorite specimen. Recalling that the antigorite specimen tested here is isotropic, it follows that the Young's modulus, $E$, and its associated attenuation, $Q_E^{-1}$ for antigorite can be directly estimated as functions of temperature and oscillation period from shear modulus and associated attenuation data, respectively, as
\begin{linenomath*}
  \begin{equation}
    E(T,T_0)=\frac{9K(T)G(T,T_0)}{3K(T)+G(T,T_0)},
    \end{equation}
\end{linenomath*}
and, for complex elastic moduli with small imaginary components \citep{WinklerNur1979},
\begin{linenomath*}
  \begin{equation}
    Q_E^{-1}=\frac{E(T,T_0)}{3G(T,T_0)}Q_G^{-1},
    \end{equation}
\end{linenomath*}
where $K$ denotes bulk modulus. Note that the temperature-dependence of antigorite's bulk modulus can be directly calculated from that experimentally measured for shear modulus (see below) by taking a temperature-independent Poisson's ratio (an assumption which seems reasonable for antigorite \citep{Christensen1996,Jietal2013}) and using conversions between elastic constants.

\section{Discussion and Conclusions}
\label{sec:disc}
Our laboratory dataset provides new quantitative estimates of the temperature dependence of shear modulus $G$ at seismic frequencies (\si{\milli\hertz}--\si{\hertz}) in the antigorite stability field. The shear velocity $V_{\scriptscriptstyle \mathrm{S}}$ and impedance $Z_{\scriptscriptstyle \mathrm{S}}$ are directly calculable as $V_{\scriptscriptstyle \mathrm{S}}=\sqrt{G/\rho}$ and $Z_{\scriptscriptstyle \mathrm{S}}=\sqrt{\rho G}$, respectively, where $\rho$ is the rock density.

At room temperature, the shear modulus is independent of frequency (Figure~\ref{fig:1672GQ}a), a behaviour which is common in dry rocks at small strain amplitudes \citep{Mavko1979,Jacksonetal1992}. The shear modulus at room temperature ($G \approx$ \SI{43}{\giga\pascal}) is broadly comparable with those previously measured by ultrasonic techniques at the \si{\mega\hertz} frequency under similar confining pressures on the same material ($G$ = \SI{37}{\giga\pascal}, at \SI{1.5}{\kilo\bar} \citep{Davidetal2018}), on other isotropic antigorite-rich serpentinites ($G$ = \SI{39}{\giga\pascal} and $G$ = \SI{34}{\giga\pascal}, respectively, at \SI{2}{\kilo\bar} \citep{Simmons1964,Christensen1978}), or using aggregate averages from single-crystal elasticity data ($G$ = \SI{38.5}{\giga\pascal}, at room pressure \citep{Bezacieretal2010}).

Relatively few data on the \emph{temperature-dependence} of shear modulus (and, more generally, of seismic velocities) in antigorite are available. \citet{Christensen1979} reports measurements of the compressional wave velocity $V_{\scriptscriptstyle \mathrm{P}}$ (at \SI{2}{\kilo\bar}) and obtains $\partial V_{\scriptscriptstyle \mathrm{P}} / \partial T$ = \SI{-0.68e-3}{\kilo\metre\per\second\per\kelvin} in the \num{25}--\SI{300}{\celsius} temperature range, a value that has often been used to account for the effect of temperature on seismic velocity in antigorite in the geophysical literature (\textit{e.g.}, \citet{CarlsonMiller2003,MookherjeeCapitani2011}). However, we note that the data of \citet{Christensen1979} were obtained on a chrysotile-rich (98$\%$) serpentinite sample. The temperature dependence of P- and S-wave velocities can also be inferred from the data of \citet{Kernetal1997}. Although Kern's data were obtained on a highly anisotropic antigorite serpentinite, the temperature dependence of the shear velocity is overall independent of the direction of both wave propagation and wave polarisation and averages \SI{-0.4}{\kilo\meter\per\second\per\celsius} in the \num{20}--\SI{620}{\celsius} temperature range (at \SI{1}{\kilo\bar}). In comparison, a linear fit to the shear modulus \textit{vs.} temperature data of Figure~\ref{fig:1672GQ}a gives $\partial G / \partial T \sim$ \SI{-0.017}{\giga\pascal\per\kelvin} (at \SI{1}{\second} oscillation period). This estimate is more robust than the value of $\partial G / \partial T$ for the Burgers model (Table~\ref{tab:burg}), which is associated with a significant uncertainty as the model was only fitted to the \num{500}--\SI{550}{\celsius} data. Differentiation of the relation $V_{\scriptscriptstyle \mathrm{S}}=\sqrt{G/\rho}$ with respect to temperature allows the temperature dependence of the shear velocity to be calculated from that of shear modulus as:
\begin{linenomath*}
  \begin{equation}
    \partial V_{\scriptscriptstyle \mathrm{S}} / \partial T=\frac{\partial G / \partial T}{2\sqrt{\rho G}}+\frac{\alpha_{\scriptscriptstyle \mathrm{V}}V_{\scriptscriptstyle \mathrm{S}}}{2},
    \end{equation}
\end{linenomath*}
where $\alpha_{\scriptscriptstyle \mathrm{V}}$ is the volumetric coefficient of thermal expansion ($\alpha_{\scriptscriptstyle \mathrm{V}} \approx$~\SI{2.8e-5}{\per\kelvin} for antigorite \citep{HollandPowell1998}). This calculation yields $\partial V_{\scriptscriptstyle \mathrm{S}}/\partial T$ = \SI{-0.76e-3}{\kilo\metre\per\second\per\kelvin}. This ``low-frequency'' temperature derivative of the shear velocity is higher than that inferred from the \si{\mega\hertz} measurements of \citet{Kernetal1997} --- a result which is expected in a dispersive material. Finally, note that, in the \num{20}--\SI{620}{\celsius} temperature range, the laboratory measurements of \citet{Kernetal1997} also provide direct estimates of $\partial V_{\scriptscriptstyle \mathrm{P}}/\partial T$ averaging \SI{-0.6}{\kilo\meter\per\second\per\celsius} (at \SI{1}{\kilo\bar}). Such a temperature derivative is very close to that measured by \citet{Christensen1979} on chrysotile serpentinite, which may indicate that the temperature-dependence of seismic velocities may be comparable among serpentine polytypes. 

It is plausible that the time dependence of attenuation above \SI{500}{\celsius} might originate from anelastic processes acting at grain boundaries, \textit{i.e.}, from intergranular relaxation. This interpretation is first supported by the overall fine-grained nature of the material, along with its pronounced grain size and shape heterogeneity. This complex grain boundary morphology, combined with the strong elastic anistropy of antigorite crystals \citep{Bezacieretal2010}, will inevitably contribute to the stress concentrations on grain boundaries that are involved in elastically accommodated and diffusionally assisted grain-boundary sliding \citep{Jacksonetal2014}, captured here in the Burgers model by the wide distribution of anelastic relaxation times. The operation of an intragranular mechanism is also unlikely due to the difficulty in activating dislocation slip systems in antigorite without very high differentail stresses \citep{Auzendeetal2015}. Furthermore, laboratory observations suggest that dislocation damping in silicates (\textit{e.g.}, in olivine) is usually only operative at temperatures significantly above the antigorite stability field \citep{Gueguenetal1989,Farlaetal2012}. In addition, the value of activation energy (\SI{268}{\kilo\joule\per\mole}) obtained in this study is an order of magnitude higher than those of \citet{Hilairetetal2007} for moderate to large strain dislocation-based deformation in antigorite at high confining pressures $\geq$ \SI{10}{\kilo\bar} (9 to 60~\si{\kilo\joule\per\mole} depending on flow law and pressure range), although large uncertainties are attached to the values of activation energy in both studies. Nevertheless, we also suggest that some viscoelastic relaxation might be concentrated within the antigorite grains, involving dislocation slip within the (001) ``corrugated'' cleavage plane, or possibly along weak conjugate planes making a small angle to the basal (001) plane and formed by weak OH bonds located between tetrahedral and octahedral sheets \citep{Amiguetetal2014}. Considering the high difficulty of identifying the relaxation mechanisms responsible for the observed anelasticity at the microscopic level, a more definitive mechanistic interpretation would notably require further experimental studies looking for a grain-size dependence, and knowledge of activation energy for elementary mechanisms such as grain boundary diffusion or self-diffusion. Although no evidence of dehydration is observed in the antigorite specimen recovered after the experiment, the present experimental conditions (confining pressure of \SI{2}{\kilo\bar}, maximum temperature of \SI{550}{\celsius}) are at or beyond the thermodynamic stability field of antigorite \citep{UlmerTrommsdorff1995}. Therefore, regardless of the specific mechanism responsible for anelastic relaxation, it is possible that the noticeable increase of attenuation above \SI{500}{\celsius} is caused by enhanced diffusion kinetics near the dehydration temperature in metastable antigorite.

The geophysical significance of the present study is highlighted by comparing the new shear modulus and attenuation data in antigorite (Figure~\ref{fig:1672GQ}) with existing data in the same temperature range in \emph{olivine}, the most abundant mineral in the Earth's upper mantle. The data of  \citet{JacksonFaul2010} and \citet{Jacksonetal2014} reveal that olivine is essentially elastic in its behaviour at temperatures lower than \SI{800}{\celsius}, with $G>$~\SI{60}{\giga\pascal} and $\log_{10} Q_G^{-1} \ll -2.0$. The contrasting seismic properties between antigorite and olivine can be summarised as follows. (i) Antigorite's shear modulus is approximately 50--60$\%$ of that of olivine at all temperatures. For instance, at \SI{400}{\celsius} and \SI{10}{\second} oscillation period, the shear moduli for antigorite and olivine are \num{37.7} and \SI{65}{\giga\pascal} (\citet{Jacksonetal2014}, Fig. 4), respectively. Taking densities of \num{3.32e3} and \SI{2.65e3}{\kilogram\per\cubic\metre} yields S-wave velocities of \num{3.77} and \SI{4.42}{\kilo\meter\per\second}, respectively. The resulting contrasts in shear velocity and impedance are substantial: about 17$\%$ and 47$\%$. (ii) The shear modulus is clearly more temperature-dependent in antigorite than in olivine. At \SI{1}{\second} oscillation period, the shear modulus of antigorite decreases by about 21$\%$ between room temperature and \SI{550}{\celsius}, whereas the corresponding decrease in the shear modulus of olivine is only about 13$\%$. (iii) Attenuation is higher in antigorite than in olivine, \textit{e.g.}, at \SI{550}{\celsius} and \SI{100}{\second} oscillation period, $\log_{10} Q_G^{-1} \approx -1.5$ for antigorite, whereas for olivine $\log_{10} Q_G^{-1} \ll -2.0$ below \SI{800}{\celsius}. A similar observation holds if dissipation data for antigorite are compared with measured values of $\log_{10} Q_G^{-1} \approx -2.0$ on a $\sim$80$\%$ olivine-rich dunite under comparable conditions (\citet{Jacksonetal1992}, Fig. 5d).

Our study reports new laboratory measurements of seismic velocity and attenuation in antigorite serpentinite at various temperatures in laboratory conditions as close as possible to the propagation of seismic waves in the lithosphere. It is noteworthy that the $Q^{-1}$ here measured in antigorite is comparable to those inferred seismologically in the mantle wedge-subducting slab region in subduction zones \citep[see \textit{e.g.}][]{P09}. The contrast in shear velocity between antigorite and olivine is significant and consistent with the commonly accepted view that the presence of serpentinite is associated with velocities that are lower than in peridotites. Attenuation is higher in antigorite than in olivine, but such contrast is much less than the shear velocity contrast. 

%% Enter Figures and Tables near as possible to where they are first mentioned:

% \begin{figure}
% \end{figure}

%\begin{figure}[ht!]
% \centering
% when using pdflatex, use pdf file:
% \includegraphics[width=20pc]{figsamp.pdf}
% If you don't specify the file extension,
%\caption{Short caption}
%\label{figone}
%\end{figure}

%\begin{figure}[ht!]
%\centerline{\includegraphics[angle=45,height=2in]{figsamp}}
%\caption{The figure caption should begin with an overall descriptive
%statement of the figure followed by additional text. They should be
%immediately after each figure.  Figure parts are indicated with
%lower-case letters ({\bf a, b, c}\ldots).  For initial submission, please place
%both the figures and captions in the text near where they are cited.}
%\label{fig:mean_and_slope}
%\end{figure}

%%% End of body of article

%%%%%%%%%%%%%%%%%%%%%%%%%%%%%%%%
%% Optional Appendix goes here
%
% \appendix resets counters and redefines section heads
% but doesn't print anything.
% After typing \appendix
%
%\section{Here Is Appendix Title}
% will show
% A: Here Is Appendix Title
%
%\appendix
%\section{Here is a sample appendix}
%This is an Appendix section.

%\subsection{subsection}
%This is an Appendix subsection.

\begin{acknowledgments}
The UK Natural Environment Research Council supported this work through grants NE/K009656/1 to NB and NE/M016471/1 to NB and TMM. This work has been enriched by discussions and help from Thomas Mitchell (UCL), and discussions with Greg Hirth (Brown University). Hayden Miller and Harri Kokkonen (ANU) provided technical support. The help of Christopher Cline (ANU) during experimental work and of Jackie Kendrick (University of Liverpool) during specimen preparation has been greatly appreciated. Ana Ferreira (UCL) provided useful comments on the manuscript. Experimental data are available from the UK National Geoscience Data Centre (http://www.bgs.ac.uk/services/ngdc/) or upon request to the corresponding author.
\end{acknowledgments}

%%  REFERENCE LIST AND TEXT CITATIONS
%

%\bibliography{mybibfile}
% 2. Run BiBTeX on your LaTeX file.
%
% 3. Open the new .bbl file containing the reference list and
%   copy all the contents into your LaTeX file here.

% 4. Run LaTeX on your new file which will produce the citations.
%
% AGU does not want a .bib or a .bbl file. Please copy in the contents of your .bbl file here.

\renewcommand\thefigure{A.\arabic{figure}}    
\setcounter{figure}{0}    
%\section{Supplementary figures}

\begin{figure*}[t]
  \centering
\includegraphics[width=\textwidth]{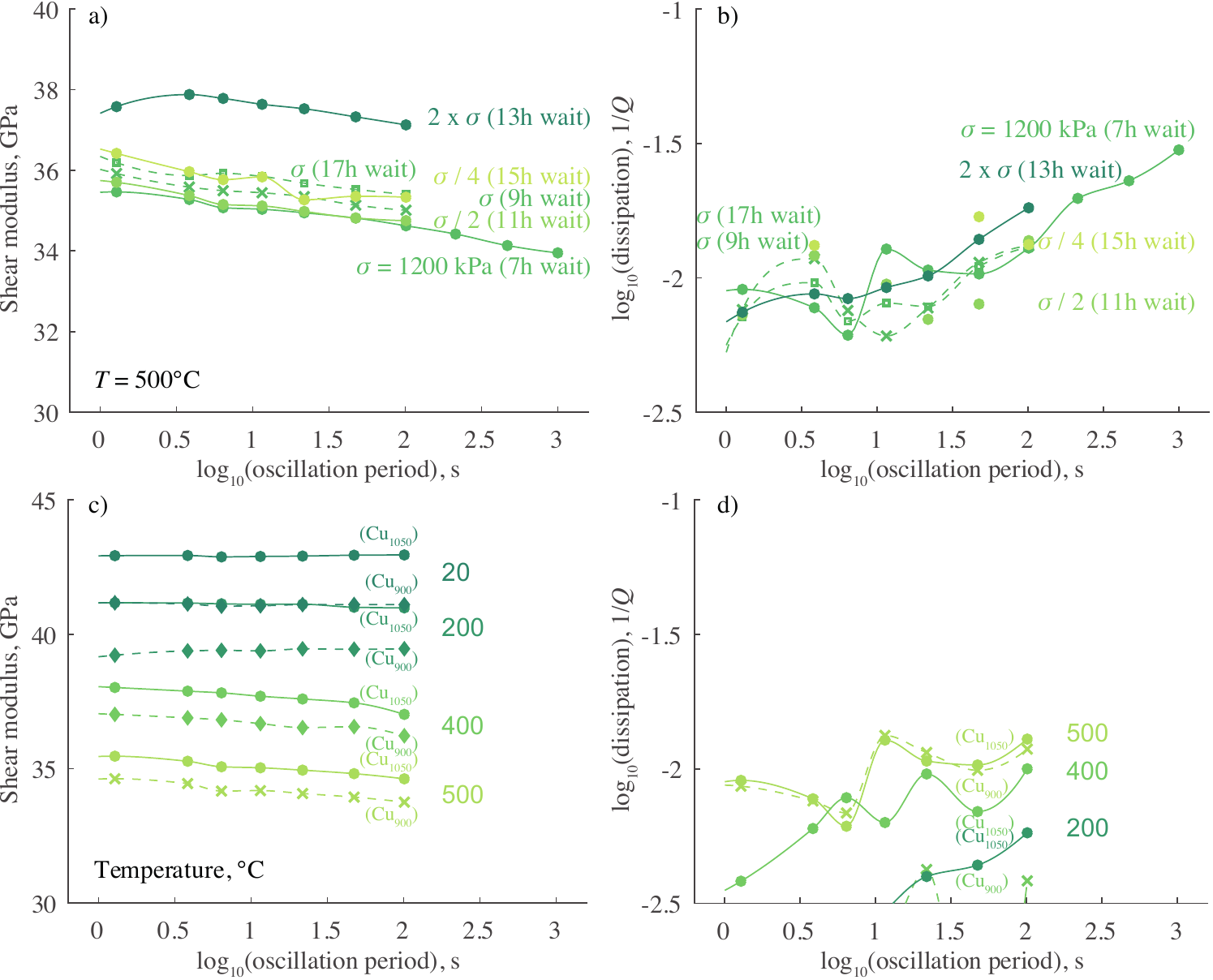}
\caption{a,b) Shear modulus and dissipation data for antigorite for various forced-oscillation tests at \SI{500}{\celsius}, documenting temporal evolution and stress amplitude dependence of the mechanical behaviour with increasing exposure. The indicated stresses are maximum values pertaining to the cylindrical surface of the specimen. c,d) Sensitivity of shear modulus and dissipation data to corrections for the viscoelasticity of the copper jacket enclosing the antigorite specimen, at selected temperatures. The viscoelastic properties of copper after annealing at \SI{900}{\celsius} (Cu$_{\textrm{900}}$) and \SI{1050}{\celsius} (Cu$_{\textrm{1050}}$) were taken from \citet{DavidJackson2018} (note that the viscoelastic properties of Cu$_{\textrm{900}}$ at \SI{500}{\celsius} were calculated by interpolating data between \num{600} and \SI{400}{\celsius}).}
  \label{fig:1672GQlinsens}
\end{figure*}

\begin{figure*}[b]
  \centering
\includegraphics[width=\textwidth]{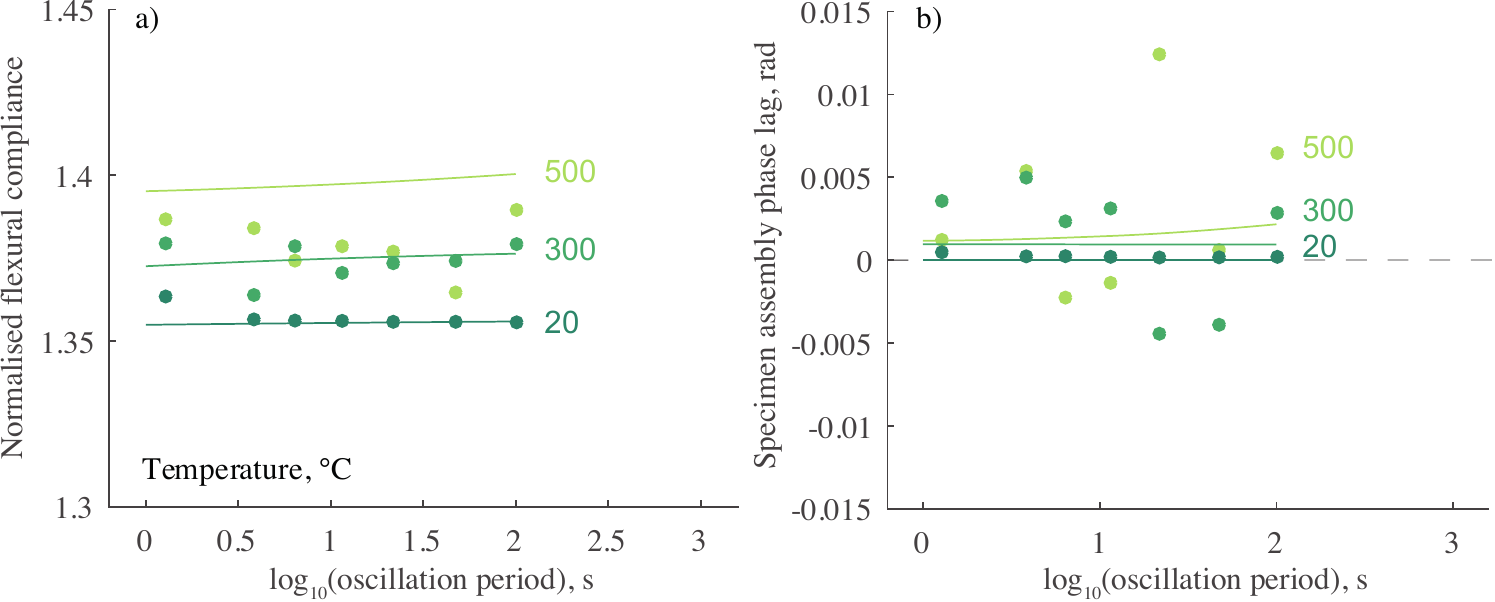}
\caption{Normalised flexural compliance and phase lag data (symbols) and flexure model (curves) for antigorite at representative temperatures.}
  \label{fig:1672flex}
\end{figure*}

\end{article}

\end{document}